# Low-cost VR Collaborative System equipped with Haptic Feedback

Samir Benbelkacem
Abdelkader Bellarbi
Nadia Zenati-Henda
Centre for Development of Advanced Technologies (CDTA), Algiers, Algeria
sbenbelkacem@cdta.dz

Ahmed Bentaleb
Ahmed Nazim Bellabaci
Institute of Electrical and Electronic Engineering (IGEE), Boumerdes, Algiers
bentalebahmed98@gmail.com

Samir Otmane
University of Evry, Paris, France
samir.otmane@ibisc.univ-evry.fr

## ABSTRACT

In this paper, we present a low-cost virtual reality (VR) collaborative system equipped with a haptic feedback sensation system. This system is composed of a Kinect sensor for bodies and gestures detection, a microcontroller and vibrators to simulate outside interactions, and smartphone powered cardboard, all of this are put into a network implemented with Unity 3D game engine.

## CCS CONCEPTS

• Interaction paradigms → Virtual reality; Collaborative interaction;
• Hardware → Sensors and actuators; Wireless devices;

## KEYWORDS

collaborative virtual reality, haptic feedback system.

## 1 INTRODUCTION

In this paper, haptic feedback system is proposed. The aim is to allow users to interact in collaborative VR environment using avatars and feel these interactions on their bodies. This haptic feedback system is implemented with vibrator motors attached to different part of the body and controlled by a microcontroller. Collisions and interactions on the virtual world are signaled to a Wi-Fi module which then sends appropriate data to the microcontroller and vibrates the motors causing a realistic sensation.

## 2 SYSTEM OVERVIEW

The system uses clients/server architecture where a Kinect is connected to a central server (Figure1).

The key aspect of our system is to allow avatars to interact between them so that their corresponding users can feel the interaction by using haptic feedback system attached on each user's body.

In our case, the users' avatars are grouped on a room and each user using, a VR box, see the avatars of other users. The Haptic feedback system receives, through Wi-Fi, a collision information that indicates which avatar and part of the avatar body collided.

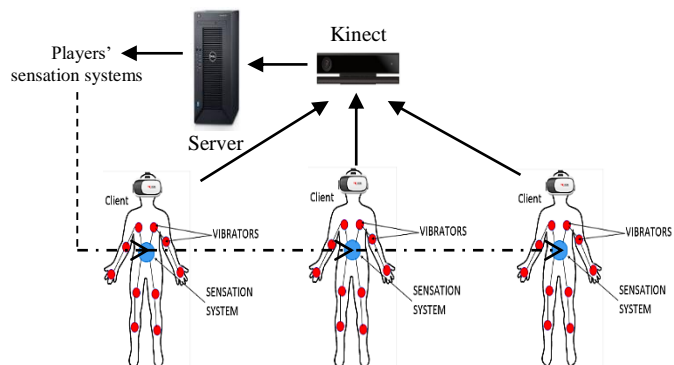

Figure 1: Low-cost VR collaborative system.

This system interprets the collision information and activates the corresponding vibrator motor located on the user's real body. This user will feel a vibration as if he has been really touched by another user.

## 3 HAPTIC FEEDBACK SYSTEM DESIGN

Two aspects have been developed. The first consists in realizing a haptic feedback while the second aspect implements a VR collaborative application using this feedback. Our haptic feedback system (called also sensation system) allows users to feel vibrations when their corresponding avatars interact. For example, we take a scene where user "A" views its avatar and the other avatars. If the user "A" touch, using its avatar, a neighbor avatar called user "B", the latter will feel a vibration at the place that corresponds to the contact of the two avatars. User "B" will feel that he has been touched by user "A".

The haptic feedback system is an electronic circuit composed of ATmega microcontroller [Reh 2013], vibrators and Wi-Fi module. The vibrators, controlled by ATmega microcontroller, are attached in different parts of users' body. The ATmega is interfaced with a Wi-Fi module (ESP-01) using esp8266 chip [Esp8266ex 2018] to connect to the network. This connection is made through TCP protocol using sockets [Fatourou 2012]. The Wi-Fi module is programmed to be as a modem or monitor (configured in station mode), so, it can act as a mini-sever. Each mini-server contains its proper IP address. The Wi-Fi module send this IP address to the central server to allow the communication between Wi-Fi module and server. Also, Wi-Fi module can be configured as client with another IP address.

The haptic feedback system evolution can be given as follows: in the beginning, the Wi-Fi module is connected to the central server and send the IP address of the station mode using TCP protocol.

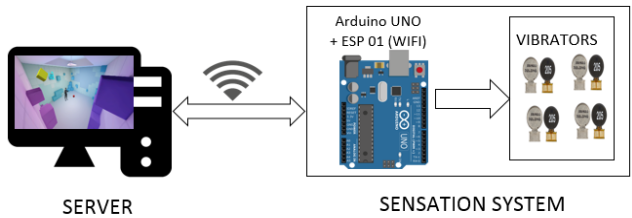

**Figure 2: Communication between haptic feedback sensation system and server.**

Wi-Fi module wait for interruption signal that corresponds to a 3D world interaction (collisions between avatars or between an avatar and a 3D object) in VR environment. When Wi-Fi module receives an interruption signal, it communicates with the microcontroller in order to actuate the corresponding vibrator engine on the human's body. Each time, an interaction (example: collision between two avatars) is made on the VR environment, the corresponding client opens a socket and sends a TCP request with its proper user ID and the body part collided to WiFi Module. The latter send these data to ATmega microcontroller through serial UART [UART 2010]. ATmega sends a signal to vibrate the corresponding motor according to the user ID and the body part collided. Figure 2 above describes the communication between the haptic feedback system and VR environment.

## 4 EXPERIMENTAL RESULTS

The USE [Davis 1989] (Usefulness, Satisfaction and Ease of Use) questionnaire was used to evaluate subjectively our work. Users were asked to rate agreement with the statements, ranging from strongly disagree to strongly agree. Users are equipped with VR headsets, haptic feedback system (electronic circuit + vibrator motors on the different parts of each user's body + Wi-Fi module), see Figure 3. We have develop a simple Unity 3D application which is composed of a red cube inserted in the middle of the scene.

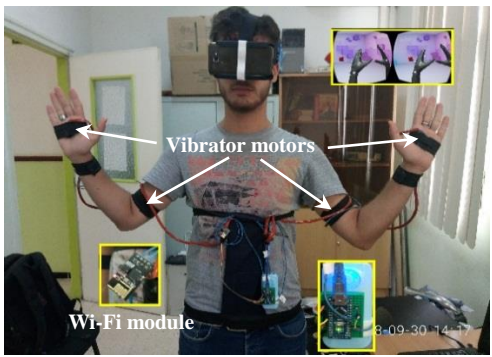

**Figure 3: User equipped with our haptic feedback system Kit.**

The users are asked to grab the cube with his hand. As soon as the collision is detected on the game, the haptic feedback system is notified. Then, it produces vibrations of the motor located on the user's hand.

A total of 20 subjects (6 females, 14 males) participated in the test. Their ages were ranged between 18 and 49. According to the filled questionnaires, most participants appreciated the simplicity and the ease of the application. They also liked the sensations and feelings provided by the system (Figure 4).

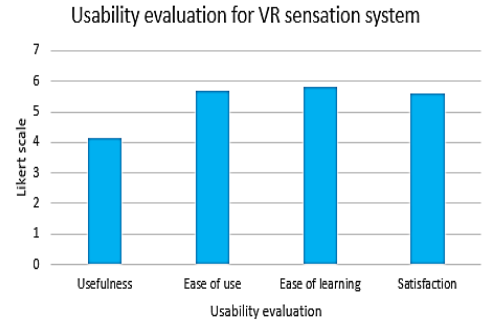

**Figure 4: Usability evaluation for VR Haptic feedback system.**

We also asked subjects to give feedback about the effectiveness of our design by answering question related to the smoothness of the interaction, realism, synchronization of motion (between VR and real world) and the 3D quality taking into account factors such as depth perception, impressiveness and visual comfort (Figure 5).

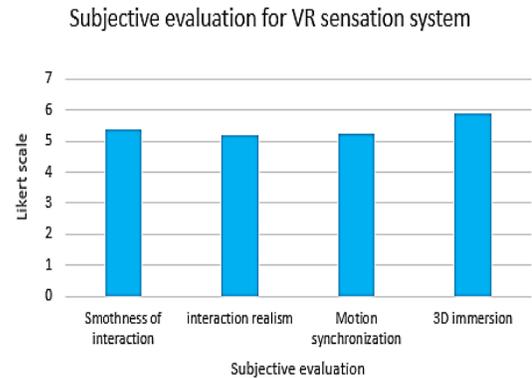

**Figure 5: Subjective evaluation for VR Haptic feedback system.**

## 5 SUMMARY

In this paper, we proposed a haptic feedback system that allow the user to feel real interactions initially produced on a VR environment. Experimental results on the proposed system show relevant results. In fact, the users participating in the evaluation had the possibility to feel their interaction initially produced on the simulation using Kinect.